# Topological nanocolloids with facile electric switching of plasmonic properties

YE YUAN[1], IVAN I. SMALYUKH[1,2,3,*]

[1]Department of Physics, University of Colorado, Boulder, CO 80309, USA
[2]Materials Science and Engineering Program, Department of Electrical, Computer and Energy Engineering, Liquid Crystal Materials Research Center, University of Colorado, Boulder, CO 80309, USA
[3]Renewable and Sustainable Energy Institute, National Renewable Energy Laboratory and University of Colorado, Boulder, CO 80309, USA
*Corresponding author: ivan.smalyukh@colorado.edu

**Combining topology and plasmonics paradigms in nanocolloidal systems may enable new means of pre-engineering desired composite material properties. Here we design and realize orientationally ordered assemblies of noble metal nanoparticles with genus-one topology and unusual long-range ordering mediated by their interactions with the surrounding nematic fluid host. Facile electric switching of these composites is reminiscent to that of pristine liquid crystals (LCs), but provides a means of reconfiguring the nanoparticle assembly and thus also the ensuing composite medium's optical properties. Our findings may lead to formation of new molecular-colloidal soft matter phases with unusual optical properties as well as optical metamaterials.**

Topological colloids [1], in which constituent particles have surface topology distinct from that of spheres, currently require sophisticated means of production such as photolithography or multi-photon photopolymerization [1-3], which may hinder their technological and fundamental-science uses. In addition to limitations with scaling of these fabrication approaches, the types of materials and the colloidal particle sizes range are also limited. Fabrication of topologically nontrivial metal or metal-alloy nanoparticles is particularly challenging, albeit wet chemistry approaches capable of controlling nanoscale composition and morphology already exist [4-7]. Topological colloids are of particular interest in the context of using liquid crystals (LCs) as host fluids [1-3, 8-12] because of the interplay between the surface topology of the constituent genus g>0 particles with the topology of the LC molecular alignment field, the so-called director field **n(r)** describing spatial patterns of rod-like LC molecular orientations. On the other hand, plasmonic nanoparticle dispersions are a class of colloids with properties very different from those of bulk materials, which are enabled by the localized surface plasmon resonances (LSPRs) and offer unprecedented means of controlling light at nanometer scales [13,14]. Dispersions of anisotropic metal nanoparticles in the LC host exhibit spontaneous fluid-host-mediated orientational ordering of the inclusions [10-12]. This ordering can be pre-engineered to exhibit both positive and negative values of the scalar order parameter of dispersed nanoparticles [10-12,15-17] and can be controlled by external fields as well as light [11]. Apart from their potential applications in nanophotonics and electro-optics, these LC-colloidal hybrid soft matter systems are of particular interest because of the richness of soft condensed matter physics behavior [18].

In this letter, we synthesize topologically nontrivial plasmonic noble metal nanoparticles with the topology of torus (genus g=1) and geometric shape of triangular frames. The sharp edges of these plasmonic nanoframes are thought to be useful for generating strong LSPRs. We explore polarization and voltage-dependent extinction spectra of this dispersion, which provide evidence for unusual orientational ordering of such frames that align, on average, at ≈45° with respect to the far-field LC director **n**₀.

Aqueous dispersions of nanoframes with a mean lateral size within the range 30-75 nm and thickness of ≈10 nm are synthesized through a two-step procedure reported previously (Fig. 1) [7], albeit it is implemented with slight modifications, as discussed below. First, silver nanoprisms of thickness ≈6 nm and lateral size 20-65 nm are synthesized using the seed-mediated method, as described in detail in ref. [19] [Fig. 1(a)]. The lateral size of these nanoprisms is controlled by varying the initial quantity of the silver seed in the synthesis within 20-500 μL [19], thus tuning the LSPRs extinction peak across the entire visible spectral range (Fig. 1). These silver nanoprisms are then used to obtain nanoframes of different lateral sizes according to a procedure we describe in detail for the particular case of 50x10 nm triangular frame-shaped particles, for which the experimental characterization of LC-nanoparticle composite is presented in this work. To synthesize them, 24.525 mL of nanoprism aqueous dispersion made from 100 μL seed is combined with an ascorbic acid solution (450μL, 10 mM), and HAuCl$_4$ solution (6 mL, 0.5 mM) added at a rate of 1 mL/min using a syringe pump (YA-12, Yale Appratus) while stirring the mixture mildly within 6 min at room temperature. This procedure results in the formation of silver-gold-alloy triangular nanoframes [Fig. 1(d)] in a process schematically depicted in Fig. 1(a). Interestingly, nanoframes with multiple holes (different large genus) are obtained at the intermediate stages of this synthesis procedure; an example of such large-genus particles obtained after 3 min is shown in Fig. 1(c). After this, 10 mL of saturated cetyltrimethylammonium bromide (CTAB) solution is added and the nanoframes are centrifuged down at 9000 rpm for 10 min to remove excess surfactant. Then, the purified silver-gold nanoframe dispersion is diluted using 40 mL of deionized water and functionalized with thiol-terminated methoxy-poly(ethylene glycol) (mPEG-SH, Jenkem Technology) according to the procedures described elsewhere [11]. To obtain nematic nanocolloidal dispersions, PEG-capped triangular nanoframes are then redispersed in 4-cyano-4'-pentilbiphenyl (5CB, Chengzhi Yonghua Display Materials Co. Ltd.) following the protocols of Ref. [11]. This method yields a composite soft matter system with colloidal nanoframes dispersed in the LC host as medium-oriented individual nanoparticles.

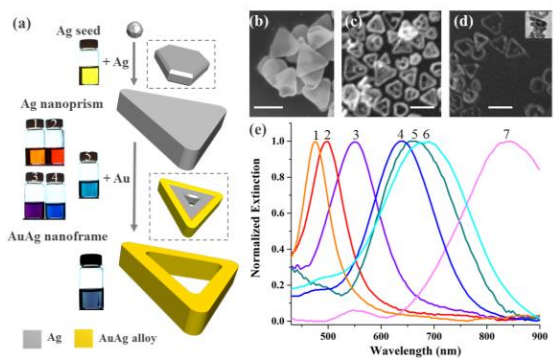

Fig. 1. Synthesis and characterization of metal nanoparticles. (a) Schematic of the synthesis procedures and structure of the nanoparticles obtained at different stages. The color of colloidal nanoparticles in aqueous dispersion pictured in the vials is indicative of the corresponding LSPR characteristics. Ag nanoprisms in the vial 5 are used to synthesize AuAg nanoframes characterized in detail in this work. (b, c, d) SEM images of (b) Ag nanoprisms, (c) AuAg alloy nanoframes with polydisperse genus obtained at the intermediate stages of the synthesis procedure, and (d) AuAg alloy nanoframes with genus g=1 obtained in the end of this procedure. Scale bars are 50 nm. The inset of (d) depicts nanoframes standing vertically on their edges, providing information on their thickness. (e) Normalized extinction spectra of as-prepared samples of colloidal Ag nanoprism in water made from different volumes of seed solution: 1) 500 μL, 2) 400 μL, 3) 280 μL, 4) 150 μL, 5) 100 μL, 6) 60 μL, 7) 20 μL. The numbers next to the spectra are correlated with the vials pictured in (a).

Planar LC cells are constructed from 0.17 or 1.0 mm-thick glass substrates coated with transparent indium tin oxide (ITO) electrodes on their inner surfaces and rubbed to impose the planar surface boundary conditions for $\mathbf{n_0}$ along the rubbing direction. The composite material is then sandwiched between two such glass substrates with the cell gap defined by silica beads of 30 μm in diameter. Homeotropic cells are constructed in a similar way, except that plain glass substrates treated in aqueous solution of dimethyloctadecyl[3-(trimethoxysilyl)propyl]ammonium chloride (DMOAP) (1 wt%) are used, defining the boundary conditions for $\mathbf{n_0}$ perpendicular to the substrates.

Polarized extinction spectra and optical images are collected using an Olympus BX51 polarizing optical microscope (POM) equipped with 10X, 20X, and 50X dry objectives (all from Olympus, numerical aperture NA=0.3-0.9), a CCD camera (Spot 14.2 Color Mosaic, Diagnostic Instruments, Inc.), and a microscope-mounted spectrometer (USB2000-FLG, Ocean Optics). Additionally, we use two other spectrometers to measure ultraviolet-visible (Evolution 60S, Thermal Scientific) and infrared spectra (HP70951A, Agilent Technologies, Inc.). To characterize spectra, the probing light first passes through the cell containing the LC-nanoframes composite material and then through a rotatable polarizer before collected by the detector. Dark field imaging and video tracking are performed using an Olympus IX81 inverted microscope, 100X oil objective with adjustable NA=0.6-1.3 (UPlanFl Olympus), dark field condenser (Olympus U-DCW, NA=1.2), and a CCD camera (Flea, PointGrey). Nanoparticle diffusion data is extracted from tracking Brownian motion of the particles by following their positions within the video frames processed by a video-tracking software ImageJ supplemented by its Wrmtrck plugin (both obtained from the NIH) [20].

Dispersion procedures similar to the ones used in the past for rods and platelets [11] readily yield LC-nanoframe composites at studied concentrations up to 1 pmol/mL in 5CB (Figs. 2 and 3). Since the nanoframes are much smaller than the surface anchoring extrapolation length ξ=K/W for 5CB-PEG interfaces, estimated to be of the order of 100-500nm [11], one can assume that the surface boundary conditions at LC-nanoparticle interfaces are weak and bulk elastic distortions are relatively small, so that their contribution to the free energy can be neglected, along with the contribution of terms associated with the gradients of the scalar order parameter. The surface anchoring energy due to LC-nanoframe interfaces, which thus plays a key role in defining behavior of this system, is expected to minimize at well-defined orientation of the nanoframes. Furthermore, since our nanoframes appear to have roughly square-shaped cross-sections [Figs. 1(d) and 2(a)], one can find the approximate value of the equilibrium orientation angle between the normal to the frame plane $\mathbf{v}$ and the far-field director $\mathbf{n_0}$ of the LC host purely from geometric considerations, without pursuing the minimization problem as we did previously [11]. The competition between surface anchoring energy costs due to the departures of the director from the easy axis orientations at the four sides of nanoframes with square cross-sections yields equilibrium orientation of $\mathbf{v}$ with respect to $\mathbf{n_0}$ at ≈45° [Fig. 2(a)]. However, the polydispersity of dimensions and detailed geometric features of the nanoframes [Fig. 1(d)] may play a role in altering this angle, which is expected to vary slightly for different particles. Additionally, since this anchoring-mediated coupling of orientations of $\mathbf{v}$ and $\mathbf{n_0}$ orientations is weak [11], thermal fluctuations also broaden the orientational distribution function. The uniaxial symmetry of the nematic host allows for a double-cone distribution of $\mathbf{v}$ at its minimum-energy orientation, i.e. at ≈45° to $\mathbf{n_0}$ [Fig. 2(b)], with the rotational Brownian motion assuring that all of these allowed low-energy orientations are present in a sample with a concentrated dispersion of the silver-gold nanoframes in 5CB [Fig. 2(c)]. This simple free energy analysis of equilibrium orientations is consistent with our experimental findings [Figs. 2(d), 2(e) and 3(d)].

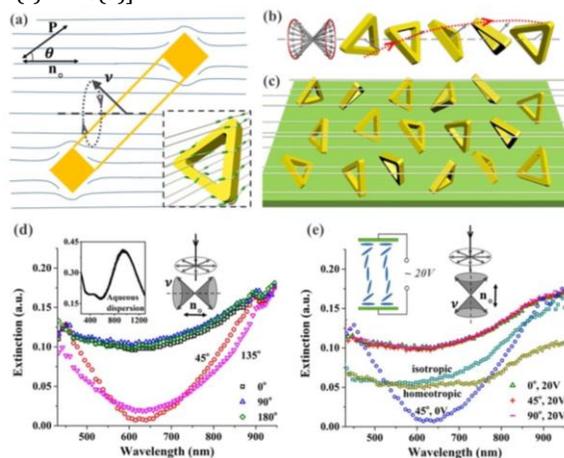

Fig. 2. Alignment of nanoframes in a nematic LC and electric switching of their LSPR spectra. (a) Cross-sectional schematic of a nanoframe with tangential surface anchoring aligned with respect to $\mathbf{n_0}$. Inset shows a three-dimensional perspective of the nanoframe and the minor local distortions of the director $\mathbf{n}$ around it. The normal to the plane of nanoframes is defined as $\mathbf{v}$. (b) A schematic showing that, considering the uniaxial nonpolar symmetry of the nematic LC host, $\mathbf{v}$ can rotate freely (driven by thermal fluctuations at no free energy cost) around $\mathbf{n_0}$, forming a double-cone. (c) Schematic of double-cone-distributed colloidal nanoframe orientations individually dispersed in the LC host. (d) Extinction spectra of the nanoframe-5CB composite in a planar LC cell. The angles between the probing polarization $\mathbf{P}$ and rubbing direction are defined in the inset of (a). The top-left inset shows the extinction spectrum of nanoframes in an aqueous dispersion. The top-right inset depicts the double-cone of $\mathbf{v}$ orientations relative to $\mathbf{n_0}$. (e) Extinction spectra of the same cell as in (d) but at applied electric field

(20V, 1kHz). Additionally, the spectra measured for this composite in isotropic phase and in a homeotropic cell are shown for comparison at zero external field. The insets schematically show how the applied field affects alignment of the LC director/molecules and thus the double-cone distribution of nanoframe orientations corresponding to the studied LSPR spectra.

The extinction spectra of silver-gold nanoframes in water shown in the top-left inset of Fig. 2(d) are qualitatively consistent with previous literature reports [21], albeit our particles are somewhat larger and our synthesis procedure was modified as compared to these previous studies in the effort to obtain approximately square cross-sections and g=1, as discussed above. When re-dispersed in 5CB, the spectra retain similar features as in water, although the higher effective refractive index of this LC dispersion medium (ranging between the 5CB's ordinary index ≈1.52 and extraordinary index ≈1.7, depending on polarization and direction of propagation the probing light) as compared to that of water (≈1.33) causes a slight red shift in spectral characteristics [Fig. 2(d) and (e)]. Since our goal is to develop new concepts for composites capable of controlling light in the visible part of optical spectrum, we focus our studies predominantly on the visible spectral range [Fig. 2(d) and (e)]. We find that the nanoframe-LC composite exhibits polarization dependence of LSPR spectra that qualitatively differs from that of all plasmonic nanoparticle-LC composites studied so far [10-12,15-17]. Unlike in the case of noble metal nanorods and nanoplatelets in the same 5CB host fluid or other LCs [10-12, 15], these LSPR spectra are found to be relatively similar at probing polarizations **P** both parallel and perpendicular to **n₀**. However, as **P** is continuously rotated with respect to **n₀** and the angle θ between them changes, we find rather significant variations of the LSPR spectra which nearly repeat themselves every 90° in θ [Fig. 2(d)], consistent with our model of orientational self-ordering of the normals **ν** to the nanoframes with respect to **n₀** [top-right inset of Fig. 2(d)].

When voltage U is applied to the cell, we observe significant changes in the spectra at U above a certain threshold ~1V [Fig. 2(e)], comparable to that of pristine 5CB and dilute nanoparticle dispersions in 5CB [10-12,15]. No significant polarization dependence is detected for the same cell at high applied fields with U ~20 V, at which the director **n** across the cell is expected to be vertical, apart from the thin cell regions very close to the confining glass substrates, as qualitatively illustrated in the top-left inset in Fig. 2(e). Homeotropic cells of the same thickness exhibit qualitatively similar polarization-independent LSPR spectra, albeit the extinction values are slightly lower due to a lower concentration of nanoframes in 5CB [Fig. 2(e)]. These LSPR spectra for planar cells at U ~20 V and for homeotropic cells are consistent with the notion that the orientations of anisotropic nanoparticles are mechanically coupled to **n** during switching [11], which implies the up-down double cone distribution of the normals **ν** of the nanoframes shown in the top-right inset of Fig. 2(e). When both the planar and the homeotropic cells are heated to the isotropic phase of 5CB (36°C and higher), the LSPR spectra change [Fig. 2(e)] because the double-cone distribution of orientations of **ν** disappears and one recovers spectra similar to the ones obtained for aqueous dispersions of nanoframes, albeit slightly red-shifted due to the higher refractive index of 5CB in the isotropic phase (≈1.59) as compared to water.

We have confirmed that the silver-gold nanoframes remain dispersed as individual nanoparticles at concentrations ranging from individual particles to the highest used concentration of 1 pmol/mL. The quality of LC alignment is not compromised by presence of these nanoparticles and the appearance and switching of the glass cells filled with 5CB-nanoframe composites is very much similar to that of pristine 5CB. Dark field microscopy [Fig. 3(c)] reveals that the frame-shaped particles undergo Brownian motion without perturbing the uniform alignment of the LC, or these distortions are too weak to detect by monitoring polarizing optical micrographs such as the ones shown in Fig. 3(a) and (b). At the same time, these glass cells exhibit polarization dependent LSPR spectra [Fig. 2(d) and (e)] and extinction showing periodic variations in the dependence on θ [Fig. 3(d)] indicating that LC host imposes the same double-cone alignment of **ν** of platelets within the entire area of the cell. Another confirmation of high quality of nanoparticle dispersion in these nanoframe-LC soft matter composites is provided by the studies of nanoparticle diffusion [Fig. 3(e)]. The values of diffusion coefficients are characterized by probing Brownian motion of the nanoframes while they are observed using dark field videomicroscopy, with a representative frame used for tracking the nanoparticle's lateral positions shown in the top-right inset of Fig. 3(e). This tracking yields Brownian motion trajectories, such as the one depicted in the top-left inset of Fig. 3(e), and the corresponding nanoparticle displacement histograms that allow us to calculate the nanoframe diffusion coefficients in directions parallel and perpendicular to **n₀** [20]. The anisotropy of diffusion coefficients stems mainly from the anisotropy of the host LC medium and the values of the diffusion coefficients are comparable to those of other noble metal nanoparticles of similar size that were studied recently in LCs [20].

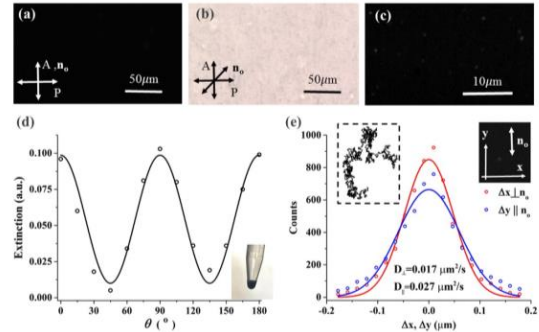

Fig. 3. Characterization of the LC-nanoframe composite and of a single particle diffusion in the LC host in a planar cell. (a, b) POM images of the composite with **n₀** at (a) 0° and (b) 45° to P of the polarizing optical microscope. (c) Dark-field image of the composite, with the bright spots corresponding to individual nanoframes undergoing Brownian motion. (d) Angular dependence of extinction at 631nm extracted from spectra similar to the ones presented in Fig. 2(d). The solid curve is for eye guiding. The inset shows the 5CB-nanoframe composite in a centrifuge tube. (e) Diffusion anisotropy of an individual nanoframe measured in directions along ($D_\parallel$) and perpendicular ($D_\perp$) to **n₀**. The top-left inset shows a representative diffusion trajectory of a nanoframe particle tracked with video microscopy. The top-right inset is dark field micrograph of a single nanoframe particle in the LC host.

Unlike micrometer-sized topological colloidal particles with genus g≥1 and with strong surface boundary conditions for **n** [1-3], which induce bulk or surface topological defects of the net hedgehog charge (or winding number for two-dimensional defects) uniquely related to g, according to the topological theorems, our g=1 nanoframes induce no defects. This very peculiar behavior stems from the weak surface boundary conditions at their surfaces and from the surface anchoring extrapolation length ξ being much larger than the nanoframe dimensions. This shows that defect-free near-uniform director structures can be obtained for topological LC colloids as well, regardless of their topological characteristics such as g, opening a large number of possibilities for designing mesostructured composites using such particles [11, 17]. In future studies, it will be of interest to explore how this behavior of topological LC colloids changes as the size of particles is continuously varied from that much smaller than ξ to much larger than ξ. Co-dispersions of plasmonic noble metal nanoframes with nanorods and nanoparticles of other geometry may provide the means of designing and practically implementing optical metamaterial behavior through exploiting molecular and colloidal self-assembly. As the volume fraction and lateral dimensions of the nanoframes can be varied through modifying the chemical synthesis procedures, one can potentially approach the concentration and nanoparticle size regimes where steric interactions between the nanoframes can pre-select a particular well-defined orientation of the normal **ν,** yielding new types of biaxial molecular-colloidal ordering in this composite soft matter system, which would be of great fundamental and practical interest. Other aspects of interest for future explorations include probing the interplay of the demonstrated double-cone orientational ordering of nanoparticles with chirality and/or partial positional ordering of the LC host fluid in cholesteric, smectic, columnar and other LC mesophases. Even in nematic LC hosts, the complex interaction of LC- colloids with external fields through different competing mechanisms, as previously observed for much larger microframes [22], may allow for unprecedented means of controlling orientational ordering of the nanoinclusions.

To conclude, we have developed a new breed of topological LC colloids with plasmonic properties and facile response to external fields. We have demonstrated that frame-shaped nanoparticles spontaneously orient at ≈45° with respect to the far-field LC director and can be effectively controlled through switching the LC director of the surrounding nematic host. This work can be extended to nanoparticles with larger genus g>1 and three-dimensional complex shapes, as well as to nanoparticles with separately controlled shape and alloy composition [4-6] and different pre-engineered surface plasmon resonance modes.

**Funding.** National Science Foundation (NSF) (DMR-1410735)
**Acknowledgment**. We acknowledge discussions with Q. Liu, A. Martinez, H. Mundoor, and Y. Zhang. We also thank A. Sanders for his help with SEM imaging and acknowledge the use of the Precision Imaging Facility at NIST, Boulder for the SEM characterization. I.I.S. is grateful for the hospitality of NREL during his sabbatical stay.